\documentclass[twocolumn,preprintnumbers,amsmath,amssymb]{revtex4}
\usepackage{graphicx}% Include figure files
\usepackage{dcolumn}% Align table columns on decimal point
\usepackage{bm}% bold math
\usepackage{hyperref} % Needed for the table
\usepackage{multirow} % Needed for the table
\usepackage{soul}
\newcommand\beq{\begin{equation}}
\newcommand\eeq{\end{equation}}
\newcommand\bea{\begin{eqnarray}}
\newcommand\eea{\end{eqnarray}}
\usepackage{wrapfig}
\newcommand{\nonum}{\nonumber}

\begin{document}

\title{\bf Modified DMRG algorithm for the zigzag spin-1/2 chain with 
frustrated antiferromagnetic exchange: Comparison with field theory at 
large $J_2/J_1$ \\}
\author{\bf Manoranjan Kumar$\bf ^1$, Zolt\'an G. Soos$\bf ^1$,
Diptiman Sen$\bf ^2$ and S. Ramasesha$\bf ^3$}
\address{\it {\rm $^1 Department$} of Chemistry, Princeton University, 
Princeton NJ 08544 \\}
\address{\it {\rm $^2 Centre$} for High Energy Physics, Indian Institute of 
Science, Bangalore 560012, India \\}
\address{\it ${\rm^3Solid}$ State and Structural Chemistry Unit, Indian 
Institute of Science, Bangalore 560012, India \\}
\date{\today}

\begin{abstract}
A modified density matrix renormalization group (DMRG) algorithm is applied 
to the zigzag spin-1/2 chain with frustrated antiferromagnetic exchange 
$J_1$, $J_2$ between first and second neighbors. The modified algorithm 
yields accurate results up to $J_2/J_1~\approx~4$ for the magnetic gap 
$\Delta$ to the lowest triplet state, the amplitude $B$ of the bond 
order wave (BOW) phase, the wavelength $\lambda$ of the spiral phase, and 
the spin correlation length $\xi$. The $J_2/J_1$ dependences of $\Delta$, 
$B$, $\lambda$ and $\xi$ provide multiple comparisons to field theories of 
the zigzag chain. The twist angle of the spiral phase and the spin structure 
factor yield additional comparisons between DMRG and field theory. Attention 
is given to the numerical accuracy required to obtain exponentially small gaps
or exponentially long correlations near a quantum phase transition.
\vskip .4 true cm
\noindent PACS numbers: 75.10.Jm, 75.10.Pq,75.40.Mg, 75.40.Cx \\ 
\noindent Email: soos@princeton.edu 
\end{abstract}

\maketitle

\section{Introduction}

Extended one-dimensional (1D) models are excellent approximations 
for the electronic structure of some crystals, either inorganic or 
organic. Quite separately, 1D models have interesting theoretical 
and thermodynamic properties. In addition to exact results, 
approximate methods have been widely applied to and tested on 1D models. 
Two major recent developments are the density matrix renormalization 
group (DMRG) and field theory. The two methods are complementary in 
principle, and both have been applied to the zigzag spin-1/2 chain 
that is the subject of this paper. In practice, however, field theory 
deals with small energy gaps or long correlations lengths near quantum 
phase transitions that may be beyond the accuracy of numerical methods, 
a point often made for Kosterlitz-Thouless transitions. The two approaches 
to extended 1D systems are quite different. DMRG is a versatile numerical 
technique for growing an extended chain from a finite one. It provides 
a complete approximate description of the ground state (gs) or other 
properties. When multiple DMRG schemes are possible, the most 
accurate one is readily identified. Field theory is an analytical 
approach based on a continuum approximation, or an effective 
Hamiltonian, to a discrete 1D model. It targets critical phenomena 
at quantum phase transitions. A 1D model may support multiple field 
theories among which it may be difficult to choose. \\

\begin{figure} 
\begin{center}
\hspace*{-0cm}{\includegraphics[width=8.5cm,height=3.0cm]{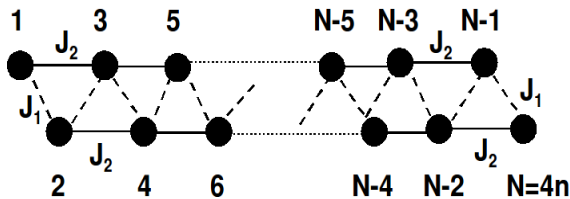}} \\
\caption{Zigzag spin-1/2 chain with AF exchange $J_1$ and $J_2$ between first 
and second neighbors, respectively. $J_1 = 0$ gives two Heisenberg chains of 
$2n$ spins with exchange $J_2$} \label{fig1}
\end{center} 
\end{figure}

In this paper, we present a modified DMRG algorithm to the zigzag 
spin-1/2 chain with frustrated antiferromagnetic (AF) exchange $J_1 > 0$ 
and $J_2 > 0$ between first and second neighbors. The Hamiltonian of 
this familiar 1D spin system is
\begin{eqnarray}
H(x)=J\sum_n ~[(1-x)\vec {S}_n \cdot \vec {S}_{n+1}+x\vec {S}_n \cdot \vec 
{S}_{n+2}]. \label{eq1} \end{eqnarray}
We consider the interval $0 \le x \le 1$ and set the total exchange 
$J = 1$ as the unit of energy. The $x = 0$ limit is a linear Heisenberg 
antiferromagnet (HAF) with many known exact properties \cite{r1} and many 
physical realizations. The $x \approx 1$ limit corresponds to two HAFs, one 
on each sublattice, and is the zigzag chain sketched in Fig. \ref{fig1}. 
Small $J_1$ for $x < 1$ or $x > 1$ 
describes an interchain exchange that is AF or ferromagnetic (F), respectively, 
and is frustrated because each spin is equally coupled to two neighbors of 
the other sublattice. The modified algorithm improves the accuracy for 
$x > 0.5$ ($J_2/J_1 > 1$).

The spin chain $H(x)$ has been extensively studied, especially in the $x = 0$ 
limit \cite{r1} that Bethe \cite{r2} and Hulthen \cite{r3} solved long ago. 
Majumdar and Ghosh (MG) found \cite{r4} a simple exact gs at 
$x_{MG}=1/3$ $(J_2 = J_1/2)$. The gs is a doubly degenerate bond order wave 
(BOW) with broken inversion symmetry at sites. The fluid-dimer transition with 
increasing $J_2/J_1$ marks the opening of a magnetic gap $\Delta(x)$ between 
the singlet gs and the lowest triplet state. The first field theoretic 
treatment \cite{r5} placed the critical ratio of $J_2/J_1$ at 1/3; subsequent 
analysis returned \cite{r6} $J_2/J_1 = 1/6$ and finally \cite{r7} $\approx 
1/4$. Okamoto and Nomura \cite{r8} obtained the accepted value, $J_2/J_1 = 
0.2411$ or $x_c = 0.1943$ in our notation, using exact results up to $N = 24$ 
sites, extrapolation and field theory. 

Classical spins in $H(x)$ lead to spiral phases for large $J_2/J_1$ when 
adjacent spins are nearly orthogonal. The gs energy per site for classical 
spins with angle $\theta$ between neighbors is
\begin{eqnarray}
E_{cl}(\theta)=J_1cos \theta+J_2cos (2\theta). \label{eq2}
\end{eqnarray}
Minimization with respect to $\theta$ leads to $cos\theta = -J_1/4J_2$ 
and $\theta = \pi/2 + \chi$ for large $J_2/J_1$. The spiral phases of quantum 
spins \cite{r9,r10,r11} are another area of interest, as are the structure 
factor \cite{r9} $S(q)$ as a function of $J_2/J_1$ and the crossover \cite{r12}
from a singlet to a ferromagnetic gs at $J_1 = -4J_2$. There are possible 
physical realizations \cite{r13} of $H(x)$, most with AF exchanges $J_1$, 
$J_2$ and a few with F exchange $J_1$.
 
White and Affleck (WA) studied \cite{r14} the BOW phase with $J_2/J_1$ beyond 
the MG point by a combination of DMRG and field theory. Numerical issues 
limited DMRG to $J_2/J_1 = 2.0$ for $\Delta$ and to 2.5 for the order 
parameter. The modified algorithm is accurate up to $J_2/J_1 = 4$. WA 
concluded that the BOW phase extends to $x = 1$ ($J_1 = 0$). Itoi and Qin 
(IQ) presented \cite{r15} a more elaborate 
field theory for large $J_2/J_1$. The present work was motivated in part by 
the contrasting results of IQ and WA. According to IQ, the spin correlation 
length diverges as \cite{r15}
\begin{eqnarray}
\xi(J_1,J_2)\approx exp(c(|J_1|/J_2)^{-2/3}), \label{eq3}
\end{eqnarray}
where $c$ is a constant. The WA expression \cite{r14} for $\xi$ has exponent 
$-1$ instead of $-2/3$ and is limited to $J_1 > 0$. The order parameter of 
the BOW phase is
\begin{eqnarray}
B(x)= \langle \vec {S}_n \cdot \vec {S}_{n+1} \rangle - \langle \vec {S}_n 
\cdot \vec {S}_{n-1} \rangle. \label{eq4}
\end{eqnarray}
$B(x)$ is the gs amplitude of the BOW for $x > x_c$. 
WA call it ``dimerization'', a term that we reserve \cite{r16} for 
structurally dimerized systems such as polyacetylene or ion-radical 
salts or spin chains. Broken inversion symmetry in a BOW phase is 
{\it electronic} dimerization in a regular array. Both WA and IQ support 
their $\xi(x)$ with the same (limited) DMRG results \cite{r14} for $B(x)$ 
and $\Delta(x)$. 

Since $B(x)$ and $\Delta(x)$ are proportional to $1/\xi(x)$, DMRG for 
the BOW amplitude or the magnetic gap can be compared to field theory as
\begin{eqnarray}
lnB(x) \approx ln\Delta(x) \approx -c(J_1/J_2)^{-2/3} \label{eq5}
\end{eqnarray}
with $J_2/J_1 = x/(1-x)$. The numerical problem is to evaluate exponentially 
small quantities at large $J_2/J_1$. The two computations are independent, 
since $\Delta(x)$ requires the triplet state while $B(x)$ does not.
DMRG directly yields approximate spin correlations functions in the gs
\begin{eqnarray}
C(p)=\langle \vec {S}_n \cdot \vec {S}_{n+p} \rangle, \label{eq6}
\end{eqnarray}
and the wavelength $\lambda(x)$ of a spiral phase, if present. As noted by 
WA \cite{r14}, $B(x)$ and $\Delta(x)$ are well-defined quantities whereas 
$\xi(x)$ requires an unknown fitting function in addition to $C(p)$. The order
parameter of the spiral phase is the twist angle $\chi(x)$ below Eq. 
(\ref{eq2}) that is related \cite{r14,r11} to the BOW phase as 
\begin{eqnarray}
\theta(x)-\frac{\pi}{2}=\chi(x)=\frac{\pi}{4 \xi}. \label{eq7}
\end{eqnarray}
$\chi(x) = 2\pi/\lambda(x)$ has been approximated by a coupled-cluster 
expansion \cite{r9} and by twisted boundary conditions in finite systems 
\cite{r11}.
 
A spiral phase of $H(x)$ has been analyzed \cite{r10} in the classical limit 
of an infinite spin at each site in terms of a nonlinear $\sigma$-model that 
involves a $3 \times 3$ orthogonal matrix. That field theory does not produce 
a BOW, however, and is not powerful enough to yield scaling results for 
$\lambda(x)$ 
or $\Delta(x)$. On the other hand, field theories \cite{r14,r15} based on 
bosonization do not predict a parameter range in which a spiral phase 
should appear. Indeed, there is no compelling field theoretic reason 
that necessarily relates the spiral and BOW phase. They have different 
order parameters and different symmetries, a discrete symmetry for translation 
by one site in the BOW phase and a continuous rotational symmetry for the 
spiral phase. The BOW extends from \cite{r8} $x_c = 0.1943$ to \cite{r14} 
$x = 1$, while the range of a spiral phase is \cite{r17} from $x_{MG} = 1/3$ 
to $x=1$. The richness of the zigzag chain at large $J_2/J_1$ makes it ideal 
for a critical discussion of DMRG accuracy and comparisons to field theory.

Section II describes the modified DMRG algorithm in which four 
rather than two spins are added per step. The accuracy improves modestly 
at $x = 0$ and dramatically for $x > 2/3$ where the second-neighbor $J_2$ 
dominates. Adding four spins when $J_1$ is small amounts to increasing 
two weakly-coupled chains by two spins each, just as adding two spins does 
at $x=0$ in conventional DMRG. We present results in Section III for $B(x)$ 
and $\Delta(x)$ up to $x = 0.8$ ($J_2/J_1 = 4$) and for $\chi(x)$ and 
$\lambda(x)$ up to $x = 0.75$ ($J_2/J_1 = 3$), the practical limit in chains 
of $N < 1000$ spins with open boundary conditions (OBC). Our results agree 
with the IQ expression in Eq. (\ref{eq3}) with $c = 2.90 \pm 0.10$ for all 
four quantities. We also compute the structure factor $S(q)$ and its maximum 
$q^*$ that yields an independent estimate of the twist angle $\chi(x)$. We 
comment in Section IV on the status of comparisons between DMRG and field 
theory for the zigzag chain at large $J_2/J_1$.

\section{Modified DMRG algorithm}

DMRG is among the most accurate numerical techniques for solving extended 
1D quantum cell models \cite{r18,r19,r20,r21}. Conventional DMRG algorithms 
start with four sites and grow an extended chain by adding two sites in 
the middle, treating the left and right half-blocks as system and environment 
by turns \cite{r18,r19}. The accuracy of this method decreases for long-range 
(beyond first neighbor) interactions since we encounter bonds between old sites
of the same system block. Long-range interactions in conventional DMRG couple 
sites at every step whose operators have undergone an unequal number of 
renormalizations. The spin chain $H(x)$ in Eq. (\ref{eq1}) has second-neighbor 
$J_2$ between sites introduced on successive steps. The decrease in accuracy 
becomes significant when $J_2$ is large because site operators involved in 
$J_2$ are renormalized twice while the $J_1$ operators are renormalized only 
once. A remedy is to add sites on every step that encompass the 
full range of interactions.
 
\begin{figure}
\begin{center}
\hspace*{-0cm}{\includegraphics[width=8.5cm,height=6.5cm]{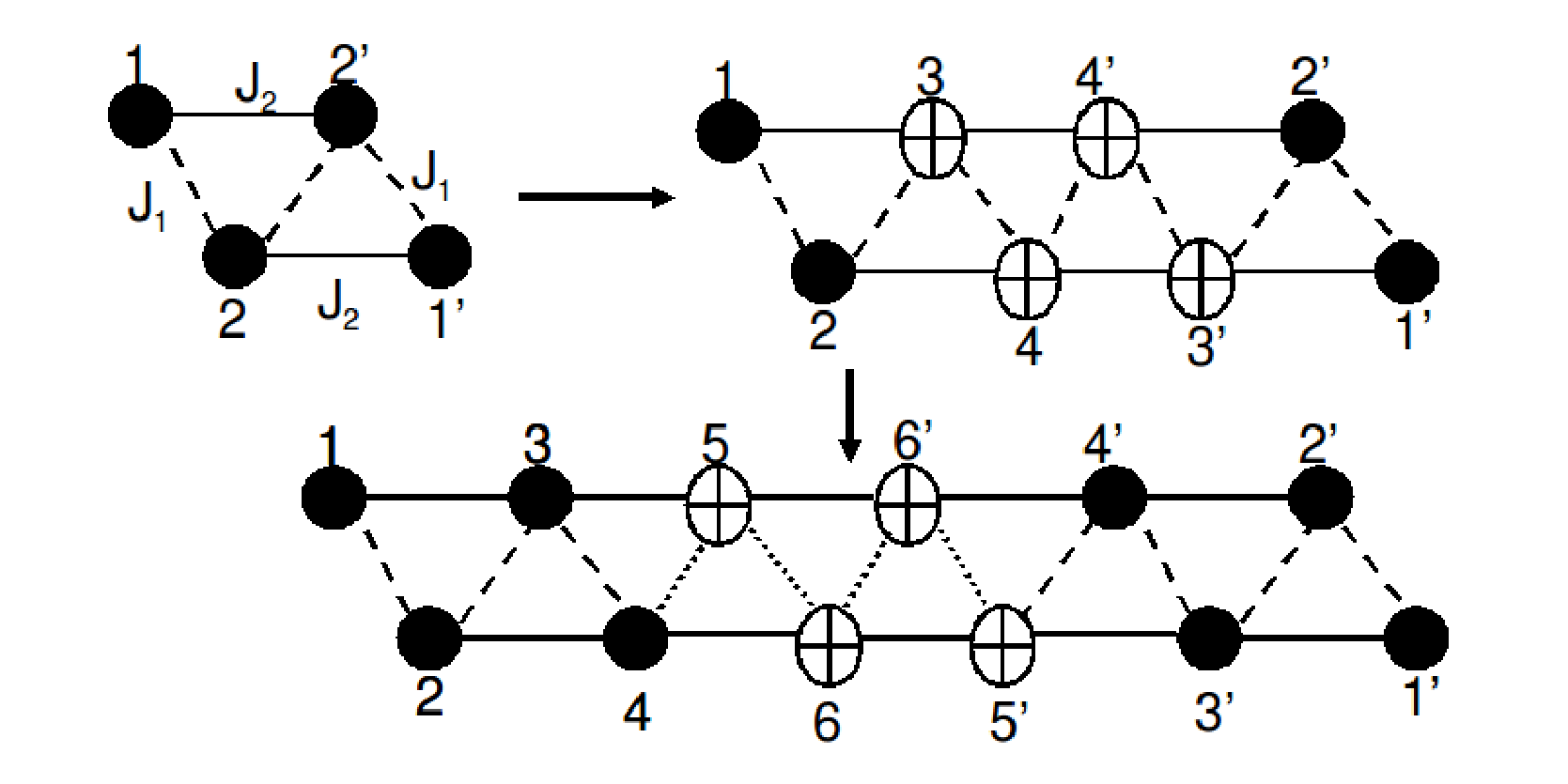}} \\
\caption{DMRG scheme with four new sites added per step. Primed and unprimed 
indices are sites of the left and right blocks, respectively. Open circles 
represent new sites and closed circles, old sites. Solid lines represent 
$J_2$, dashed lines $J_1$.} \label{fig2}
\end{center}
\end{figure}

Accordingly, we modified the DMRG algorithm for $H(x)$ to add two new sites 
per half block instead of one, as shown schematically in Fig. \ref{fig2}. The 
system starts with 4 spins and grows to $N = 4n$ in $N - 1$ steps. Large 
$J_2/J_1$ leads to weakly coupled chains of $2n$ sites in Fig. \ref{fig1}, 
each with a singlet gs when $J_1 = 0$. The Fock space dimensionality of each 
block increases as $2(2s+1)m$, or as $4m$ for $s = 1/2$ sites, which is 
comparable to fermionic systems. We find that keeping $m = 150$ eigenvectors 
of the density matrix is sufficient for good accuracy. The truncation error 
in the sum of the eigenvalues of the density matrix is less than $10^{-9}$ 
in the worst case, and increasing $m$ changes the energy only in $\rm 5^{th}$ 
or $6^{th}$ decimal place, in units of $J$. For more accurate spin correlation
functions $C(p)$ and order parameter $B(x)$, we used finite DMRG calculations 
on every fourth steps \cite{r19}. $B(x)$ is calculated using the middle bonds 
of the chain and is accurate up to 5-6 decimal place, but is 
subject to finite-size effects of order $1/N$ discussed below.

\begin{figure}
\begin{center}
\hspace*{-0cm}{\includegraphics[width=7.0cm,height=9.7cm,angle=-90]{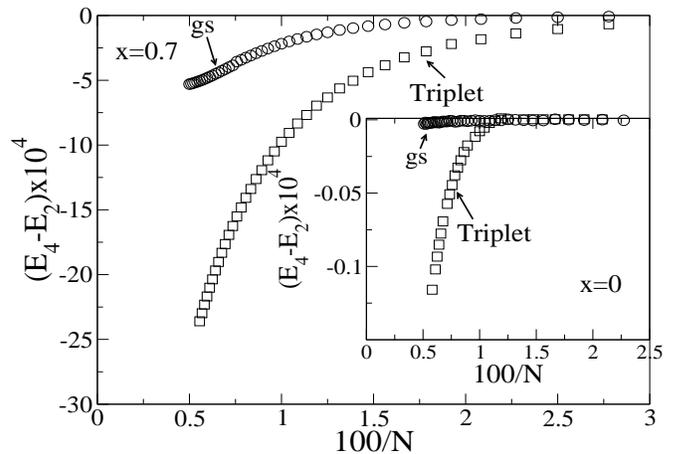}} \\
\caption{ Energy difference per site between the new ($E_4$) and conventional 
($E_2$) DMRG for the singlet gs and the lowest triplet at $x = 0.70$ and 
$x = 0$ (inset) for chains of $N$ sites.} \label{fig3}
\end{center}
\end{figure}

We compare the modified algorithm with four sites added per step to 
conventional DMRG for the gs energy and the gap magnetic $\Delta$. We use 
the infinite DMRG algorithm with $m = 200$ in each case. Since DMRG targets 
the lowest state in each $M_S$ sector, the lowest permissible total spin 
state has the best energy in an AF model. Conventional DMRG is most accurate 
at $x = 0 (J_2 = 0)$ where there is only nearest-neighbor exchange. 
Nevertheless, as shown by the inset in Fig. \ref{fig3}, the new method 
improves the gs energy slightly and the triplet energy considerably. Note 
that the inset energy scale is 100 times finer that of the main figure. We 
attribute better performance to (i) the absence of ‘old-old’ bonds within the 
same block in the new scheme, and (ii) increased number of new-new bonds 
(3 at $x = 0$) compared to new-old bonds (2 at $x = 0$) when four new sites 
are added at each step. The conventional ratio is 1:2 at $x = 0$. The accuracy
of the new method at $x = 0$ is about $10^{-7}$ for the singlet and $10^{-5}$
for the triplet. It runs smoothly for $x > 2/3$, in contrast to numerical 
difficulties \cite{r14} of conventional DMRG at $x > 1/2$. The estimated 
accuracy for $x > 0.5$ is $10^{-5}$ for the gs and $10^{-3}$ for the triplet. 
We also studied chains with $J_2 > 0$ and F exchange $J_1 < 0$ in terms of 
$J_1$, $J_2$ rather than $x>1$ in Eq. (\ref{eq1}). 

\begin{figure}
\begin{center}
\hspace*{-0cm}{\includegraphics[width=7.0cm,height=9.7cm,angle=-90]{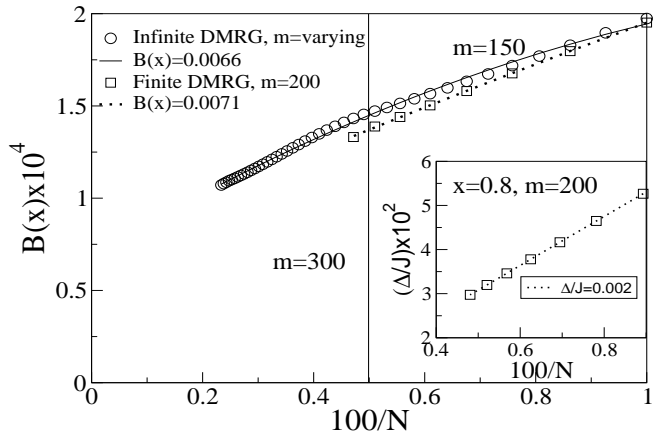}} \\
\caption{BOW order parameter $B(0.80)$ vs $1/N$ for finite and infinite DMRG 
and (inset) magnetic gap $\Delta(0.80)$ vs $1/N$. Infinite DMRG is carried out
with $m=300$ for $N>200$ and $m=150$ for $N<200$; finite DMRG has $m = 200$.}
\label{fig4}
\end{center}
\end{figure}

Figure \ref{fig4} shows the size dependence of $B(0.8)$ and $\Delta(0.8)$ at 
$J_2/J_1 = 4.0$. These are the smallest $B$ and $\Delta$ that are accurate 
with the present DMRG. We varied $m$ to look for jumps in $B(x)$, such as 
those in Fig. \ref{fig6} of ref. 14 at $J_2/J_1 = 2.5$, but found only the 
smooth behavior shown. In the four-site algorithm, $B(x)$ is linear in $1/N$ 
for large $N$. Finite DMRG procedure with $m = 200$ and $N$ between 100 and 
200 sites returns $B(0.8) = 0.0071$, as shown in Fig. \ref{fig4}. The infinite
algorithm with variable $m$, and $200 \le N \le 430$ leads to extrapolated 
$B(0.8) = 0.0066$. The inset of Fig. \ref{fig4} shows the $1/N$ dependence 
of $\Delta(0.8)$ using finite DMRG with four spins added per step. The 
extrapolated gap is $0.002$. Similar extrapolation at $x = 0$ give $\Delta 
\approx 0.001$, close to the exact $\Delta = 0$. 

\section{The $J_2/J_1> 1$ regime of $H(x)$}

\begin{figure}
\begin{center}
\hspace*{-0cm}{\includegraphics[width=7.0cm,height=9.7cm,angle=-90]{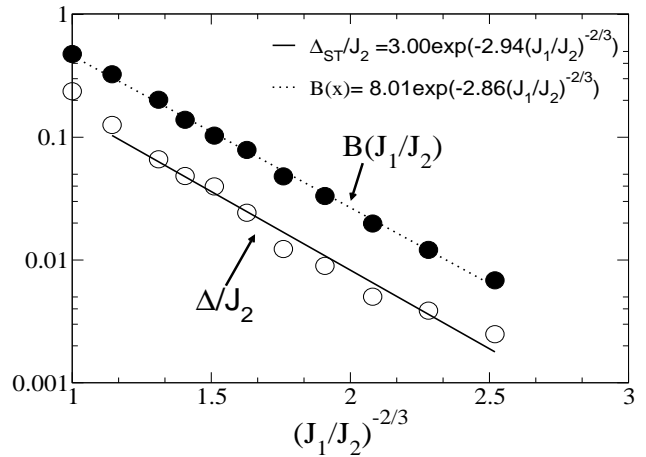}} \\
\caption{Order parameter $B$ and magnetic gap $\Delta$ as a function of 
$(J_1/J_2)^{-2/3}$. The fitted lines are $1/\xi(x)$ in Eq. (\ref{eq3}) with 
$c = 2.94$ for $\Delta$ and 2.86 for $B$.} \label{fig5}
\end{center}
\end{figure}

Larger $J_2/J_1$ is accessible with the improved DMRG algorithm. All results 
below are for $m = 200$ and OBC for $N = 800$ sites, as discussed in Section 
II. The order parameter $B(x)$ in Eq. (\ref{eq4}) and the magnetic gap 
$\Delta(x)$ from the gs to the lowest triplet provide direct comparison 
to field theory of the BOW phase. Both $B(x)$ and $\Delta(x)$ go as 
$1/\xi(x)$, where $\xi(x)$ is the correlation length in Eq. (\ref{eq3}). As 
seen in Fig. \ref{fig5}, the IQ exponent of $-2/3$ fits the DMRG results 
remarkably well up to $J_2/J_1 = 4$ $(x = 0.8)$ with $c=2.90$. The $B(x)$ 
fit covers almost two orders of magnitude and extends to $J_2 = J_1$.

The $J_2/J_1 = 1$ point for $\Delta$ deviates upward from the line in Fig. 
\ref{fig5}. IQ used DMRG \cite{r14} for $\Delta(x)$ in the interval $0.6 \le 
J_2/J_1 \le 2$ to support $\xi$ in Eq. (\ref{eq3}) with $c = 3.66$. This is 
not correct, and the one-loop approximation does not extend down to $J_2/J_1 
\approx 1$. WA used DMRG for $B(x)$ up to $J_2/J_1 = 2.5$ to support $\xi$ 
with exponent $-1$ instead of $-2/3$ in Eq. (\ref{eq3}). Their expression fails
at larger $J_2/J_1$. DMRG with $0.6 \le J_2/J_1 \le 2.5$ is not appropriate 
for the BOW phase at large $J_2/J_1$. Although Fig. \ref{fig5} covers more 
than an order of magnitude in $\Delta$ and almost two for $B$, there is no 
assurance that $J_2/J_1 = 4$ is large enough. On the other hand, if $J_2/J_1 
\approx 4$ is not ``large'', numerical comparison 
with field theory will indeed be difficult.

The accuracy of $B(x)$ is limited by finite-size effects for OBC and $N = 800$
sites. We illustrate with an uncorrelated example. A half-filled H\"uckel 
or tight-binding chain of $N$ sites has bond orders \cite{r22}
\begin{eqnarray}
p_m=2\sum^{N/2}_{k=1}c_{k,m}c_{k,m+1}, \label{eq8}
\end{eqnarray}
with $m = 1,2,...,N-1$. The coefficient $c_{k,m}$ at site $m$ of the filled 
orbital $k$ is
\begin{eqnarray}
c_{k,m}=\sqrt{\frac{2}{N+1}}sin{\frac{\pi km}{N+1}}. \label{eq9}
\end{eqnarray}
The geometrical series for $p_m$ is summed for finite $N$. The difference 
between $p_{N/2}$ of the central bond and that of either neighbor is 
$-2(-1)^{N/2}/N$ for large $N$. 
The bond order $p_{N/2}$ is less than the band limit of $2/\pi$ for $N = 4n$ 
and greater than $2/\pi$ for $N = 4n+2$, just as expected for partial 
single and double bonds at the center of linear polyenes with evenly 
spaced C atoms. Since OBC break inversion symmetry at sites, this elementary 
example has implications for any OBC simulation of BOW systems. In any case, 
the exponential decrease of $B$ with $J_2/J_1$ is soon overwhelmed by $1/N$ 
corrections that limit DMRG with $N \approx 1000$. Finite-size corrections to 
$\Delta$ or other low-energy excitations also go as $\approx 1/N$ and place 
similar limits on the accuracy of exponentially small gaps.

\begin{figure}
\begin{center}
\hspace*{-0cm}{\includegraphics[width=8.0cm,height=9.7cm,angle=-90]{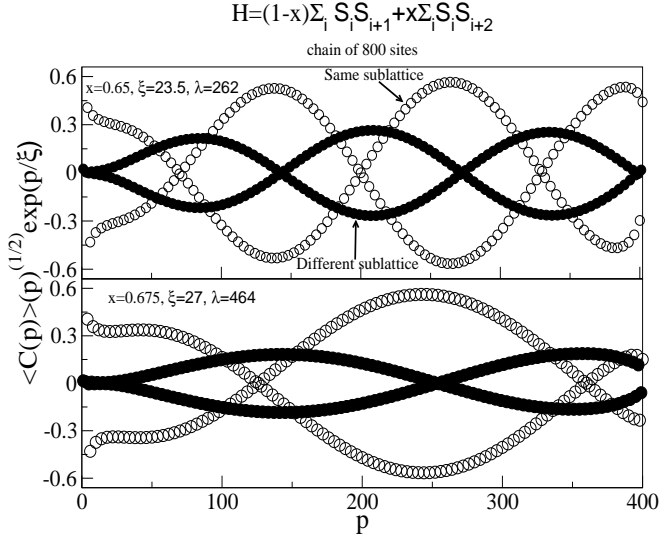}} \\
\caption{Spin correlation functions $C(p)$ on the same sublattice (open 
symbols, even $p$) and on opposite sublattices (closed symbols, odd $p$) for 
$N = 800$ sites. The scaling of $C(p)$ and choice of are discussed in text.}
\label{fig6}
\end{center}
\end{figure}

A DMRG calculation returns all gs spin correlations functions $C(p)$ in Eq. 
(\ref{eq6}). OBC implies that $C(n,p)$ depends on the site index $n$ as well as 
the separation $p$. It is customary to take sites $n$ and $n+p$ in the 
central part of the chain. $C(p)$ between sites on one sublattice in Fig. 
\ref{fig1} has even $p$, while $C(p)$ between sublattices has odd $p$. Fig. 
\ref{fig6} shows $C(p)$ in spiral phases at $x = 0.65$ in the top panel and 
$x = 0.675$ in the bottom panel. The wavelength $\lambda(x)$ of the spiral 
phase appears directly provided that there are two nodes to specify 
$\lambda/2$. DMRG with $N = 800$ sites yields $\lambda$ only up to $x = 0.75$.
The scale factor $p^{1/2}exp(p/\xi)$ follows WA, who \cite{r14} considered 
even $p$ and chose $\xi$ to make the amplitudes in Fig. \ref{fig6} as equal 
as possible. The same $\xi$ holds for odd $p$. This procedure minimally 
requires two maxima and hence is also limited to $x = 0.75 (J_2/J_1 = 3)$. 

\begin{figure}
\begin{center}
\hspace*{-0cm}{\includegraphics[width=7.0cm,height=9.7cm,angle=-90]{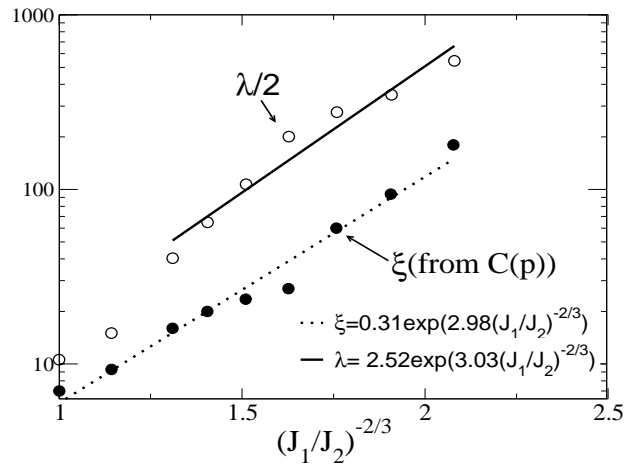}} \\
\caption{Spin correlation length $\xi$ of the BOW phase and the wavelength 
$\lambda(x)$ of the spiral phase as functions of $(J_1/J_2)^{-2/3}$, based on 
the $C(p)$ results on Fig. \ref{fig6}.} \label{fig7}
\end{center}
\end{figure}

We obtained $\lambda(x)$ and $\xi(x)$ from $C(p)$ results with $N = 800$ 
sites. DMRG is unbiased in the sense that neither a spiral nor a BOW phase is 
assumed. Fig. \ref{fig7} shows how $ln\xi(x)$ and $ln\lambda(x)$ increase 
with $J_2/J_1$. The IQ exponent of $-2/3$ in Eq. (\ref{eq3}) fits reasonably 
well over a smaller range of $J_2/J_1$ with $c = 3.03$ for $\lambda$. The 
$\xi $ exponent is consistent with the more accurate $c = 2.90$ in Fig. 
\ref{fig5} for $B(x)$ and $\Delta(x)$ over a wider range. The scaling form 
of $\xi(x)$ in the BOW phase and $\lambda(x)$ in the spiral phase are almost 
identical. According to Eq. (\ref{eq7}), the product $\lambda(x)B(x)$ or of 
$\lambda(x)\Delta(x)$ should be constant, independent of $x$ for large 
$J_2/J_1$. The calculated points in Figs. \ref{fig5} and \ref{fig7} between 
$J_2/J_1 = 1.3$ and 3.0 yield $\lambda(x)B(x) \approx 9 \pm 3$ and 
$\lambda(x)\Delta(x) \approx 5.6 \pm 2$. Since neither is monotonic in 
$J_2/J_1$, our results are weakly consistent with constant 
$\lambda(x)/\xi(x)$. Higher accuracy is needed to test Eq. (\ref{eq7}). 

The spiral phase of $H(x)$ has been modeled \cite{r9,r10,r11} in terms of the 
twist angle $\chi > 0$ for AF exchange that is defined below Eq. (\ref{eq2}) 
for classical spins. The inverse relation between $\lambda(x)$ and $\xi(x)$ 
in Eq. (\ref{eq7}) has been proposed \cite{r11,r14} for large $\lambda(x)$ or 
small $\chi$ when the discrete nature of the spin chain is irrelevant. It 
follows that
\begin{eqnarray}
C(2p)& \propto &cos(2p\theta)=(-1)^p cos(2p\chi), \nonum \\ 
C(2p+1)& \propto&cos[(2p+1)\theta]=-(-1)^pcos(2p\chi). \label{eq10}
\end{eqnarray}
Even and odd $C(p)$ are not quite out of phase in Fig. \ref{fig6}, in agreement
with Eq. (\ref{eq10}). The nodes of $C(2r)$ occur at $2r\chi=(n+ 1/2)\pi$, 
while those of $C(2r+1)$ are at $(2r + 1)\chi = n\pi$. The angle $\chi(x)$ 
decreases with increasing $J_2/J_1$ and has been studied by other techniques 
\cite{r9,r11}. Independent evaluation of $\chi(x) = 2\pi/\lambda(x)$ provides 
a consistency check for direct DMRG results for $\lambda(x)$ in Fig. 
\ref{fig6}. Such consistency is different from the common scaling of $\xi(x)$ 
and $\lambda(x)$ discussed above.

\begin{figure}
\begin{center}
\hspace*{-0cm}{\includegraphics[width=7.0cm,height=9.7cm,angle=-90]{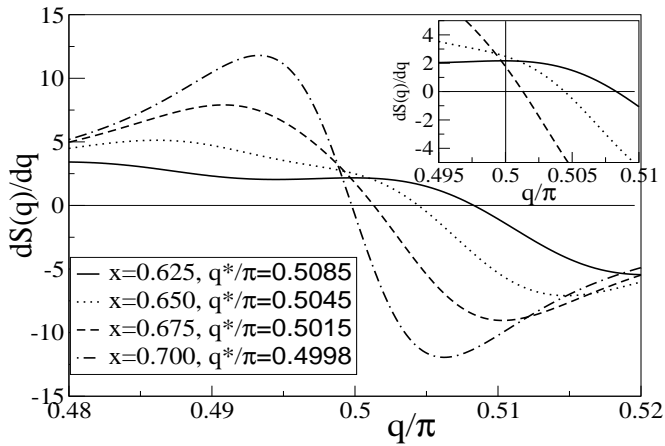}} \\
\caption{Derivative $S'(q)$ of the structure factor. The inset shows $S’(q^*) 
= 0$ for $x = 0.625$, 0.65 and 0.675. The twist angle $\chi$ is $q^* -\pi/2$ 
in radians. The $x=0.700$ result is unphysical.} \label{fig8}
\end {center}
\end{figure}

Aligia et al. \cite{r11} obtained $\chi(x)$ using twisted boundary conditions 
in Eq. (\ref{eq1}) and exact results up to $N = 24$. Bursill et al. \cite{r9} 
presented several approximation schemes for $\chi(x)$, one of which is based 
on the peak $q^*$ of the structure factor $S(q)$. The spin-1/2 structure factor
for a system with periodic boundary conditions is
\begin{eqnarray}
S(q)=\frac{1}{N} \sum_{np}C(p){\rm exp}(iqp)=\frac{3}{4}+\sum_{p=1}2C(p)
cos(qp). \label{eq11}
\end{eqnarray}
where $C(p)$ are spin correlation functions in Eq. (\ref{eq6}). Inversion 
symmetry is restored in a BOW phase by taking a linear combination of the 
degenerate gs. The MG point at $x = 1/3$ has short-range correlations, known 
exactly, leading to $S_{MG}(q) = 3(1-cos q)/4$ and a broad maximum at $q^* = 
\pi$. The maximum value $S(q^*)$ is obtained using the derivative 
\begin{eqnarray}
\frac{\partial S(q)}{\partial q}=\sum_{p=1} 2pC(p)sin(qp). \label{eq12}
\end{eqnarray}
Eq. (\ref{eq12}) shows that $q^*$ is sensitive to long-range spin correlation 
functions. We again use $C(p)$ from the central part of the chain. $C(p)$ 
refers to $n=N/2=400$ in Eq. (\ref{eq6}) and the sum is from $p=1$ to $N/2-10$, 
or 10 sites from chain end. The resulting $S'(q)$ are shown in Fig. 
\ref{fig8}. The inset magnifies the $S'(q^*) = 0$ region for the indicated 
values of $x$. As $\xi(x)$ increases and correlations become long ranged, 
large $p$ must be retained in the sum and the inherent $1/N$ limitations of 
OBC are again encountered. Although $q^* = \pi/2 + \chi(x) \rightarrow \pi/2$ 
with increasing $x$ as expected, the condition $S'(q^*)= 0$ has 
limited value in the crucial region of small $\chi$. The point 
$q^* = \pi/2$ occurs at $J_1 = 0$ that separates the AF regime with $J_1 > 0$ 
and $q^* > \pi/2$ from the F regime with $J_1 < 0$ and $q^* < \pi/2$. We 
underestimate $q^*$ for $x=0.70$, which is clearly unphysical, and hence 
overestimate $\lambda$ based on $S(q)$, but the twist angle and wavelength 
are consistent for $x < 0.65$. 

Aligia et al. \cite{r11} emphasize that twisted boundary conditions extend 
$\chi(x)$ to much larger $J_2/J_1 \approx 30$. They report reasonable 
agreement with WA \cite{r14} and with Bursill et al. \cite{r9} up to 
$J_2/J_1 = 2.5$, where our results are similar. But $ \lambda(0.8) \approx 
1300$ estimated from $J_2/J_1 = 4$ in their Fig. 4 is about 7 times smaller 
than the extrapolation of $\lambda(x)$ in Fig. \ref{fig7}. Moreover, their 
\cite{r11} asymptotic regime starts at $J_2/J_1 = 15$ where their $\xi(x)$ 
has the WA form with exponent $-1$ in Eq. (\ref{eq3}). The stronger presumed 
decrease of $1/\xi(x)$ in the spiral phase as $x \rightarrow 1$ would lose 
out to the weaker singularity of the BOW phase. Twisted boundary conditions 
up to $N \approx 24$ do not to give reliable \cite{r11} $\Delta(x)$, however, 
and no $B(x)$ results were presented. 
 
\section{Discussion}

We obtained more accurate results for the frustrated spin chain $H(x)$ 
in Eq. (\ref{eq1}) with $J_2/J_1 >1$ by modifying the DMRG algorithm to 
add four sites per step instead of two. The order parameter $B(x)$ in 
Eq. (\ref{eq4}) is limited by $1/N$ corrections in systems with open 
boundary conditions (OBC). The accuracy of the magnetic gap $\Delta(x)$ 
to the lowest triplet is estimated by comparison to exact results in 
the fluid phase with $x < x_c = 0.1943$. As seen in Fig. \ref{fig5},
 we find an exponential decrease of $B(x)$ and $\Delta(x)$ up to 
$J_2/J_1 = 4$ that follows the IQ \cite{r15} correlation function 
in Eq. (\ref{eq3}) for almost two decades. DMRG automatically yields 
the spin correlation functions $C(p)$ in Eq. (\ref{eq6}) and a spiral 
phase with wavelength $\lambda(x)$ in Fig. \ref{fig6}. Following 
WA \cite{r14}, the correlation length $\xi(x)$ is extracted from 
amplitudes in the spiral phase. Exponentially increasing $\lambda(x)$ 
and $\xi(x)$ in Fig. \ref{fig7} again follows the IQ expression in 
Eq. (\ref{eq3}), albeit over a narrower range up to $x = 0.75 (J_2/J_1 = 3)$ 
set by numerical considerations. The maximum $q^*$ of the structure factor in 
Fig. \ref{fig8} is an independent estimate of twist angle $\chi(x) = 
2\pi/\lambda(x)$ of the spiral phase in Eq. (\ref{eq2}). We find that $q^*$ 
has limited accuracy for our $C(p)$ for $x > 0.65$. 

Detailed comparison with theory is made possible by multiple studies of the 
BOW \cite{r14,r15} and spiral \cite{r9,r11} phases of $H(x)$ with $J_2/J_1 > 
1$. More generally, we wondered whether DMRG is capable of confirming the 
small gaps or long correlation lengths predicted by field theory. Our results 
to $J_2/J_1 = 4$ clearly favor the IQ expression \cite{r15} for $\xi(x)$ in Eq. (\ref{eq3}) while just as 
clearly ruling out their fit \cite{r15} for $\Delta(x)$. Greater accuracy is 
needed for meaningful comparisons. The modified algorithm yields multiple and 
reasonably consistent comparisons up to $J_2/J_1 = 4$.

The modified algorithm runs smoothly for $J_2 > 0$ and $J_1 < 0$. 
The IQ expression for $\xi(x)$ in Eq. (\ref{eq3}) does not 
depend on the sign of $J_1$. We find finite gaps $\Delta$ on the F side that, 
however, are less than our estimated numerical accuracy. Still higher accuracy 
is needed for $\Delta$ on the F side. We can definitely say, however, that the 
constant $c \approx 2.9$ for $\xi(x)$ on the AF side is different from that on 
the F side. In view of small $\Delta$, Itoi and Qin discuss \cite{r15} 
the spin wave velocity of the singlet or triplet and present 
conventional DMRG results for $N\Delta$ vs. $1/N$ in Figs. \ref{fig3} 
and \ref{fig4} of Ref. \onlinecite{r15}. The appearance of a nonsinglet 
gs at $J_1 \approx -2J_2$ contradicts the exact result of Dmitriev 
et al. \cite{r12}, that the singlet/ferromagnetic phase boundary 
of the zigzag chain is at $J_1 = -4J_2$. Field theory on the F side 
is numerically untested so far.

DMRG accounts naturally for coexisting BOW and spiral phases 
with onsets at $x_c = 0.1943$ and $x_{MG} = 1/3$, respectively, but cannot 
say where they terminate. Bosonization field theories \cite{r14,r15} 
have a BOW phase but not a spiral phase, while field theory \cite{r10} 
or other approaches \cite{r9,r11} to the spiral phase do not yield a 
BOW. Eq. (\ref{eq7}) is an assumed \cite{r14,r11} relation between 
the twist angle $\chi(x)$ of the spiral phase and the correlation 
length $\xi(x)$ of the BOW phase. The scaling of $\lambda(x)$ and $\xi(x)$ 
in Fig. \ref{fig7} is almost the same, and the products $\lambda(x)B(x)$ 
and $\lambda(x)\Delta(x)$ are roughly constant, but greater accuracy is 
needed to confirm that $\lambda(x)$ and $\xi(x)$ are indeed proportional. 
It may be interesting in the future to study whether similar scaling is 
special to spin-1/2 or holds also for higher spin.

Field theory is a continuum approximation. Since solid-state models are 
discrete, field theory becomes accurate when $\xi(x)$ exceeds 5-10 lattice 
constants. This is well documented for solitons in the SSH model \cite{r23} 
and its continuum version \cite{r24}. As shown in Fig. \ref{fig7}, $\xi(x) > 
10$ requires $J_2/J_1 > 1$ and our DMRG extends to $\xi \approx 300$. We do 
not consider the discreteness of the lattice to be important. 

It is a well-recognized numerical challenge, to obtain exponentially small 
energy gaps or exponentially long correlation lengths near a quantum phase 
transition. Impressive gains in numerical accuracy are required to be modestly
closer to the critical point. The modified DMRG algorithm for $H(x)$ extends 
accurate results to $J_2/J_1 = 4$ and clearly favors the correlation function 
$\xi(x)$ in Eq. (\ref{eq3}) proposed by Itoi and Qin \cite{r15}. There are 
open questions such as whether $J_2/J_1 = 4$ is in the 
asymptotic limit or the relation between BOW and spiral phases. Convincing 
comparison between field theory and numerical methods are in fact demanding as 
we have illustrated for the zigzag spin-1/2 chain.

{\bf Acknowledgments.} We gratefully acknowledge partial support for work at 
Princeton by the National Science Foundation under the MRSEC program 
(DMR-0819860). SR thanks DST India for funding through SR/S1/IC-08/2008 and 
JC Bose fellowship.

\end{document}